\documentclass[a4paper,11pt]{article}
\pdfoutput=1 

\usepackage{jcappub} 

\usepackage{latexsym,amssymb,amstext,amsmath}
\usepackage{mathrsfs}
\usepackage{enumerate}
\usepackage{mathtools}

\title{\boldmath {On the Trans-Planckian Censorship Conjecture and the Generalized Non-Minimal Coupling}}

\author{Omer Guleryuz}
\affiliation{Department of Physics, Istanbul Technical University,\\Maslak 34469 Istanbul, Turkey}

\emailAdd{omerguleryuz@itu.edu.tr}

\abstract{We investigate the Trans-Planckian Censorship Conjecture (TCC) and the arising bounds on the inflationary cosmology caused by that conjecture. In that investigation, we analyze TCC bounds for both Jordan and Einstein frames in the presence of a generic non-minimal coupling (to gravity) term. That term allows us to use the functional freedom it brings to the inflationary Lagrangian as an effective Planck mass. In this sense, we argue one should consider the initial field value of the effective Planck mass for the TCC. We show that as a result, one can remove the TCC upper bounds without the need to produce a new process or go beyond the standard inflation mechanism, with the generalized non-minimal coupling, and for Higgs-like symmetry-breaking potentials.}

\begin{document}
\maketitle
\flushbottom

	\section{Introduction}{\label{Intro}}
		
One can consider the observed cosmology of the early universe with an exponential acceleration scenario, known as inflation theory \cite{Guth:1980zm,Starobinsky:1980te,Linde:1981mu,Albrecht:1982wi}. That scenario is favored by the observational data from the cosmic microwave background (CMB) measurements \cite{Perlmutter:1998np,Aghanim:2018eyx,Akrami:2018odb,Percival:2009xn}. As these CMB measurements become more precise, the consistency of the minimally coupled single-field slow-roll inflationary models is weakened. In particular, most of them are ruled out by comparison with the observational parameters as spectral index $n_s$ and the tensor-to-scalar ratio $r$. (See e.g., \cite{Martin:2013tda} for the first Encyclopaedia Inflationaris.) However, one can also treat gravity as an effective quantum field theory \cite{Donoghue:1994dn}. In this sense, the predictions of these minimally coupled scenarios can vary, and be consistent with the existence of a non-minimal coupling (to gravity) term \cite{Abbott:1981rg,Amendola:1990nn,Fakir:1990eg,Faraoni:1996rf,Futamase:1987ua,Lucchin:1985ip,Salopek:1988qh,Spokoiny:1984bd,Faraoni:2004pi,Ashoorioon:2019kcy}. Embedding the Higgs boson of the standard model with a non-minimal coupling term as a part of the inflation mechanism is one of the first exciting applications of this term \cite{Bezrukov:2007ep,Barvinsky:2009fy,Bezrukov:2008ej,Clark:2009dc,DeSimone:2008ei}. The non-minimal coupling term also provides a renormalizable scalar field theory in curved spacetime and quantum corrections to the scalar field \cite{Buchbinder:1992rb,Callan:1970ze,Freedman:1974ze,Allen:1983dg,Batalin:1984jr}.

Another important consequence of the inflation theory is that it establishes a bridge between the quantum fluctuations in the early universe and the classical fluctuations that can be observed in the late universe. However, it indicates that the evolution of sub-Planck quantum fluctuations has become classical and frozen. That corresponds to the classical observations of sub-Planckian quantum fluctuations and, known as the trans-Planckian problem \cite{Brandenberger:2000wr,Brandenberger:2012aj,Easther:2002xe,Kaloper:2002cs,Martin:2000xs}. In this sense, the recently proposed Trans-Planckian Censorship Conjecture (TCC) \cite{Bedroya:2019snp} states that; the modes of sub-Planckian quantum fluctuations should never trans the Planck length, even though the fluctuations we observe today are classical. That statement mathematically corresponds to the inequality 
\begin{equation}
    a_{f}  H_{f}<\frac{a_{i}}{l_{\mathrm{pl}}} \simeq a_{i} M_{\mathrm{pl}}.
\end{equation}
Thus, any quantum gravity candidate that passes through the conjecture is in the string landscape, else, in the swampland. This inequality sets upper bounds on some variables and parameters of the considered theory. In particular, these upper bounds are immensely tight for minimally coupled single-field slow-roll inflationary models \cite{Bedroya:2019tba}. In that matter, many logical and sound ideas have been followed and analyzed \cite{Kadota:2019dol,Schmitz:2019uti,Brandenberger:2019eni,Kamali:2019gzr,Burgess:2020nec,Sanna:2021kik,Andriot:2020lea,Saito:2019tkc,Cai:2019dzj}.\footnote{Even though this is not the approach of our work, it is worth mentioning that; these bounds get more flexible for non-standard scenarios like multi-stage inflation ($\Delta H \neq 0$), or non-standard thermal history scenarios \cite{Tenkanen:2019wsd,Berera:2019zdd,Brahma:2019unn,Das:2019hto,Goswami:2019ehb,Dhuria:2019oyf,Kamali:2019xnt,Mizuno:2019bxy,Torabian:2019zms,Shi:2020ymp,Cai:2019igo}.} However, so far, considering the standard inflation paradigm and its approaches, it has not been possible to relax these upper bounds sufficiently. 

In this study, by looking at the same problem with a generalized non-minimal coupling perspective (that provides a universal attractor behavior for inflation at a strong coupling limit \cite{Kallosh:2013tua}), we aim to show that TCC bounds can be eliminated (or relaxed) naturally. In that process, we take advantage of the Higgs-like symmetry-breaking potentials as an example.

The organization of this work is as follows. In section \ref{sec2}, we first briefly talk about the minimal inflationary Lagrangian and the corresponding TCC upper bounds. Subsequently, we introduce the non-minimal coupling setup with a generic coupling term, with two different approaches in the following subsections. We also analyze both approaches for the TCC bounds. We discuss the motivations of Jordan frame TCC in section \ref{sec3}. Following that, in \ref{sec3.1} we present a case where these bounds are removed (or relaxed) with a generic non-minimal coupling term, in a model correlated way. In \ref{sec3.2}, we denote the superconformal embedding of the corresponding model. Finally, we present our concluding remarks in section \ref{sec4}.

\section{TCC and Non-Minimal Coupling}\label{sec2}
Inflationary Lagrangian traditionally is considered as a combination of the gravitational sector and the matter sector. It includes the Einstein-Hilbert term $\sqrt{-g} R$ to generate the gravitational part of the General Relativity (GR), together with a matter Lagrangian in analogy with a point-like particle Lagrangian $m (\Dot{x})^2/2 - V(x)$, to produce the matter part of the GR. For the canonical single-field inflation case, the minimal Lagrangian is given as
\begin{equation}\label{mL1}
    \mathcal{L}_{\text{minimal}} =\sqrt{-g} \left[\frac{M_{\mathrm{pl}}^2}{2} R - \frac{1}{2}\left(\partial \phi \right)^2 - V(\phi)\right]
\end{equation}
where $\phi$ is the inflation field. This Lagrangian is in the "Einstein" frame by definition since its gravitation part is the standard Einstein-Hilbert Lagrangian. The (reduced) Planck mass can be given in terms of the gravitational constant with
\begin{equation}
    M_{\mathrm{pl}} = \frac{1}{\sqrt{8 \pi G}} \simeq 2,4 \times 10^{18} \; \text{GeV} \quad \rightarrow \quad G=\frac{1}{8\pi M_{\mathrm{pl}}^2}.
\end{equation}
For this case, the slow-roll parameter $\epsilon$ takes the usual form as
\begin{equation}\label{epsilonminimal}
    \epsilon_{\text{minimal}}(\phi)=\frac{M^{2}_{\mathrm{pl}}}{2}\left(\frac{V^{\prime}(\phi)}{V(\phi)}\right)^{2}
\end{equation}
where prime means derivative with respect to the variable of the function. The Trans-Planckian Censorship Conjecture \cite{Bedroya:2019snp} for the minimally coupled case can be denoted as
\begin{equation}
    a_{f}  H_{f} < a_{i} M_{\mathrm{pl}}.
\end{equation}
As a result, given in ref. \cite{Bedroya:2019tba}, it sets an upper bound on the scalar potential which, eventually leads to an upper bound on the power spectrum of gravitational waves as 
\begin{equation}\label{Vminimal}
    V \lesssim 10^{-40}  M_{\mathrm{pl}}^4 \quad \rightarrow \quad \mathcal{P}_h(k) \sim \frac{H^2}{M_{\mathrm{pl}}^2} \lesssim 10^{-40}
\end{equation}
which, severally constraints the scale of the inflationary observable $r$ as
\begin{equation}
\mathcal{P}_{\mathcal{R}}(k)=\frac{1}{8 \pi^{2} \epsilon_{\text{minimal}}}\left(\frac{H(k)}{M_{\mathrm{pl}}}\right)^{2} \quad \rightarrow \quad r=16 \epsilon_{\text{minimal}} < 10^{-30}.
\end{equation}
Here, it is assumed that the quantum fluctuations of the inflaton are responsible for the origin of the early universe, and $\mathcal{P}_{\mathcal{R}}(k) \sim 10^{-9}$ is taken as the observed value of the power spectrum of curvature fluctuations \cite{MUKHANOV1992203,Aghanim:2018eyx}. These bounds are calculated under the assumptions of $H_{f}-H_{i} = \Delta H \simeq 0$ during inflation, and $ H_{R} = H_{f} \simeq H_{i} =$ for the instant reheating after inflation.
\subsection{Modified Gravitational Sector (Jordan Frame)}
On the other hand, the scalar field $\phi$ can contribute to the inflationary process by a non-minimal coupling term in the gravitational part as
\begin{equation}\label{JL1}
    \mathcal{L}_J =\sqrt{-g} \left[\frac{F(\phi)}{2} R - \frac{1}{2}\left(\partial \phi \right)^2 - V_{J}(\phi)\right]
\end{equation}
where $F(\phi)$ is a generic function of $\phi$ and is called the non-minimal coupling term. Since the usual Einstein-Hilbert term is deformed for this case, this Lagrangian is called in as the "Jordan" frame. Here, one can decompose the non-minimal coupling term into two main parts:
\begin{equation}
    F(\phi) = M_{\mathrm{pl}}^2 + \xi f(\phi)
\end{equation}
where $f(\phi)$ is a generic function of $\phi$, and $\xi$ is a coupling constant. The first part is for to obtain the usual GR term and the second part is for to modify gravity. As one can see, for $\xi=0$ or $f(\phi)=0$ this Lagrangian becomes the minimal Lagrangian given in (\ref{mL1}). In this frame, one can consider the non-minimal coupling term to act as the effective Planck mass via the relation $\mathcal{M}_{\mathrm{pl}}(\phi)^2 \equiv F(\phi)   $. Then the effective gravitational constant becomes:
\begin{equation}\label{Geff}
    G_{\text{eff}} =\frac{1}{8\pi \mathcal{M}_{\mathrm{pl}}^2}.
\end{equation}
Here, the non-minimal coupling term needs to be under consideration in two different ways according to the potential $V(\phi)$ is associated with symmetry-breaking solutions via a non-zero vacuum expectation value $v$ or not. For example, it is appropriate to take $f(\phi) = \phi^2$ in order to be compatible with the renormalization counter-term \cite{Buchbinder:1992rb,Callan:1970ze,Freedman:1974ze}. In the presence of a symmetry-breaking potential; one can include the non-zero vacuum expectation value into the picture by taking $f(\phi) = \phi^2 -v^2$ as first suggested in ref. \cite{Bostan:2018evz}. This term requires a symmetry-breaking  ($v=\phi$) after the inflation to restore the usual Einstein-Hilbert term $\sqrt{-g} R/2$. If we stick to the modified gravitational sector approach; since for this case the Planck mass becomes the effective Planck mass $ M_{\mathrm{pl}}  \rightarrow  \mathcal{M}_{\mathrm{pl}}(\phi) = \sqrt{F(\phi)}$, the TCC becomes:
\begin{equation}\label{NMTCC}
    a_{f}  H_{f} < a_{i} \mathcal{M}_{\mathrm{pl}}(\phi).
\end{equation}
Then, the constraint on the scalar potential in the non-minimal perspective becomes
\begin{equation}
    V_{J} \lesssim 10^{-40}  \mathcal{M}_{\mathrm{pl}}^4
\end{equation}
where the Jordan frame potential can be denoted as $V_{J} \simeq 
 3  H^{2} \mathcal{M}_{\mathrm{pl}}^2$ under the slow-roll approach. The upper bound on the power spectrum of the gravitational waves stays the same
\begin{equation}
    \mathcal{P}_h(k) \sim \frac{H^2}{\mathcal{M}_{\mathrm{pl}}^2} \lesssim 10^{-40}
\end{equation}
with the effective Planck mass $M_{\mathrm{pl}} \rightarrow \mathcal{M}_{\mathrm{pl}}$, therefore the bound on the inflationary parameter $r$ remains unchanged as
\begin{equation}
\epsilon_{J} \sim \frac{10^{9}}{8 \pi^{2}}\left(\frac{H}{\mathcal{M}_{\mathrm{pl}}}\right)^{2} \simeq \frac{10^{9}}{24 \pi^{2}} \frac{V_{J}}{\mathcal{M}_{\mathrm{pl}}^4} \quad \rightarrow \quad r=16 \epsilon_{J} < 10^{-30}.
\end{equation}
 Even though the upper bound on the $r$ remains the same, the upper bound on the potential $V_J(\phi)$ can change for different values of the effective Planck mass $\mathcal{M}_{\mathrm{pl}}(\phi)$.

\subsection{Modified Matter Sector (Einstein Frame)}
So far, we covered the non-minimal coupling term in the gravitational sector with an effective Planck mass. It is a well-known method to divert the non-minimal coupling term into the matter sector to switch back to the Einstein frame (see \cite{Kaiser:2010ps}, for an explicit derivation), to be able to use the familiar formulas of GR and inflationary parameters. The method is as follows: First, one needs to define a conformal transformation of the metric as 
\begin{equation}
    g^{\mu \nu} = \hat{g}^{\mu \nu} \frac{F(\phi)}{M_{\mathrm{pl}}^2} 
\end{equation}
which, leads to the Einstein frame Lagrangian
\begin{equation}
    \mathcal{L}_{E} =\sqrt{-\hat{g}} \left[\frac{M_{\mathrm{pl}}^2}{2} \hat{R} - \frac{Z(\phi)^{-1}}{2}  \left(\hat{\partial} \phi \right)^2 - \hat{V}_{E}(\phi)\right]
\end{equation}
where
\begin{equation}\label{VE}
   Z(\phi)^{-1}:=M_{\mathrm{pl}}^2 \left[\frac{1}{ F(\phi)} +\frac{3}{2 }\left(\log(F(\phi))\right)^{\prime 2}\right],  \quad \text{and} \quad \hat{V}_{E}(\phi) = \left(\frac{M_{\mathrm{pl}}^4 }{F(\phi)^2}\right) V_{J}(\phi)
\end{equation}
is the relation between the Jordan frame potential and the Einstein frame potential. Then, after applying a field redefinition to restore the canonical kinetic term as
\begin{equation}
\left(\frac{d \varphi}{d \phi}\right)^2 = \frac{1}{Z(\phi)},
\end{equation}
the Jordan frame Lagrangian (\ref{JL1}) changes into an Einstein frame Lagrangian with canonical kinetic terms:
\begin{equation}\label{EL1}
\mathcal{L}_{E} =\sqrt{-\hat{g}} \left[\frac{M_{\mathrm{pl}}^2}{2} \hat{R} - \frac{1}{2}\left(\hat{\partial} \varphi \right)^2 - \hat{V}_{E}(\varphi(\phi))\right].
\end{equation}
As one can see, the Jordan frame non-minimally coupled Lagrangian (\ref{JL1}) is equivalent to the Einstein frame minimally coupled Lagrangian (\ref{EL1}) after a conformal transformation and a field redefinition. This new canonical fields lead to the slow-roll parameter $\epsilon$ as
\begin{equation}
    \epsilon_{E}(\varphi) = \frac{M_{\mathrm{pl}}^2}{2} \left( \frac{\hat{V}^{\prime}_{E}(\varphi)}{\hat{V}_{E}(\varphi)}\right)^2
\end{equation}
with the usual form. As we continue,  it is possible to acquire the same constraint on the Jordan frame potential via the modified matter sector approach by inserting the equation (\ref{VE}) for the minimal case (\ref{Vminimal});
\begin{equation}\label{TCCVJ}
 V = \hat{V}_{E} \lesssim 10^{-40}  M_{\mathrm{pl}}^4 \quad \rightarrow \quad V_{J}(\phi) \lesssim 10^{-40}  F(\phi)^2,
\end{equation}
since the potential is in the Einstein frame for the minimally coupled case by definition. Using the same method, the constraint on $r$ reads
\begin{equation}\label{rEbound}
\epsilon_{E} \sim \frac{10^{9}}{8 \pi^{2}}\left(\frac{H}{M_{\mathrm{pl}}}\right)^{2} \simeq \frac{10^{9}}{24 \pi^{2}} \frac{\hat{V}_{E}}{M_{\mathrm{pl}}^4} = \frac{10^{9}}{24 \pi^{2}} \frac{V_{J}}{F(\phi)^2} \quad \rightarrow \quad r=16 \epsilon_{E} < 10^{-30}
\end{equation}
as expected. Even though both frames predict the same upper bound on the power spectrum of the gravitational waves, the modified matter sector approach is more illuminating to see the different boundaries on the Einstein frame potential and the Jordan frame potential.

\section{Jordan Frame and Model Dependence}\label{sec3}
The physical equivalence of the Einstein frame and Jordan frame for physical predictions has long been debated \cite{Faraoni:2006fx,Domenech:2016yxd,Bahamonde:2016wmz,Bahamonde:2017kbs,Bhadra:2006rn,Bhattacharya:2017pqc,Brooker:2016oqa,Capozziello:1996xg,Capozziello:2010sc,Chiba:2013mha,Corda:2010ye,Dick:1998ke,Domenech:2015qoa,Faraoni:1999hp,Faraoni:1998qx,Flanagan:2004bz,Nozari:2009ds,Postma:2014vaa,Qiu:2012ia,Qiu:2012np,Qiu:2014apa,Quiros:2012rnn,Karam:2017zno,Jarv:2016sow,Burns:2016ric,Kuusk:2016rso,Karam:2018squ,Karamitsos:2018lur,Mukherjee:2021otm}. Naturally, with a conformal transformation of the metric, all length scales can be rescaled, and one can relate and interpret both frames as equivalent (on the classical level). That means; for any given Lagrangian in Einstein frame, there is an infinite family of different Jordan frames that describe the same theory. Consequently, the conformal equivalence of these two frames allows one to generalize the mathematical formulation of TCC as
\begin{equation}
\vspace{0.1cm}
a_{f}  H_{f} < a_{i} M_{\mathrm{pl}} \; \ \xrightleftharpoons[\text{\bf Einstein Frame}]{\text{\bf \,Jordan Frame\,}} \  \;  a_{f}  H_{f} < a_{i} \mathcal{M}_{\mathrm{pl}}(\phi)
\end{equation}
for a more comprehensive description. (This formulation of the TCC is so far model-independent.) Then, it is a valid question to ask if there is any physical motivation to consider working in the Jordan frame instead of the Einstein frame. One can approach this ambiguity in two ways.
\begin{enumerate}
    \item[\textbf{i.}] {\textbf{Theory approach:}}
    Given the phenomenological and observational achievements of inflation theory, it is natural to expect that it also has its origin from a more fundamental theory of quantum gravity. Following this expectation, one can embed inflation theory into a string theoretical setup with a non-minimal coupling term in its gravity part since one also expects the presence of non-minimal couplings in string theory. One should also note here that the physical quantities of inflation theory, such as the power spectrum of the curvature fluctuations and gravitational waves, remain unaltered for both frames. They are frame-independent.
     \item[\textbf{ii.}] {\textbf{Conjecture approach:}}
     To understand the logic behind the Jordan frame TCC approach, firstly, one should reconsider the original TCC formulated in the Einstein frame. In the Einstein frame, the theory has Planck mass $M_{\mathrm{pl}}$, and therefore TCC is related to Planck length $l_{\mathrm{pl}}$, the smallest meaningful physical length in the effective field theory. However, in the Jordan frame, effective Planck mass is dynamic and directly related to the non-minimal coupling term. That means this effective Planck mass can take much higher or lower values depending on the theory itself. This indefiniteness leads to the freedom to work on different mass (or length) scales such as string, compactification, or grand unification theory scales, which appear more naturally in Jordan frames upon compactification and are more desirable for string cosmology \cite{Alvarez:1984ee,Brandenberger:1988aj,Tseytlin:1991xk,Gasperini:1992em,Lidsey:1999mc,Gasperini:2002bn,Barrow:1998wd,Brandenberger:2000itq,Easson:2000mj}. As an example, the fundamental length scale of the string theory is string length $l_{\mathrm{s}}$, which relates to string mass as $M_{\mathrm{s}} \propto l_{\mathrm{s}}^{-1}$. Using the same logic, one can easily relate the Jordan frame TCC conjecture to string scale by choosing the effective Planck mass accordingly, as $\mathcal{M}_{\mathrm{pl}}  \propto M_{\mathrm{s}}$.
\end{enumerate}
In the following subsection, we will examine a class of Jordan frames strongly correlated with the model in question.

\subsection{Removing the TCC bound with Generalized Non-Minimal Coupling}\label{sec3.1}
\begin{figure}[htb]
\centering
\includegraphics[width=4.0in]{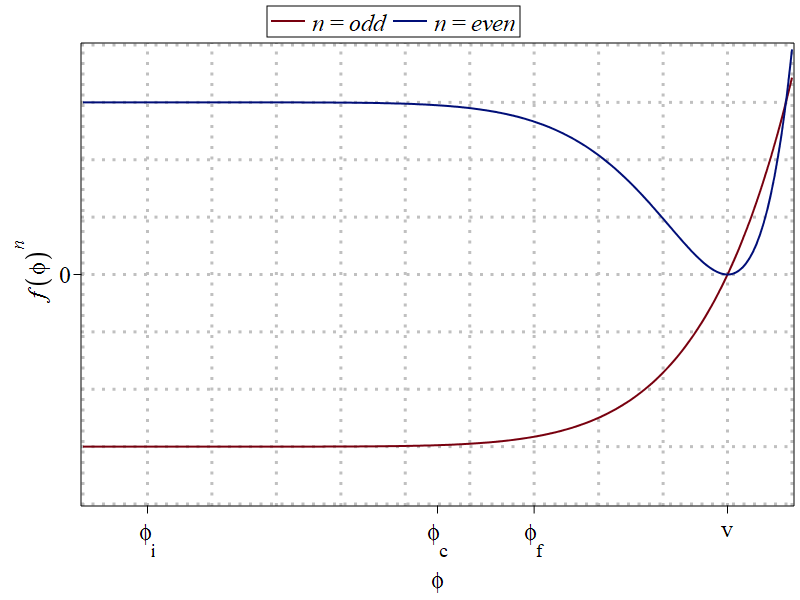}
\caption{\label{F1} A sketch of the function $f(\phi)^n$ for $f(\phi)  \propto (\phi^m - v^m)$ where $m$ is any positive integer. Here, $\phi_i$ and $\phi_f$ corresponds to the initial and final field values during the inflation. $v$ is the vacuum expectation value and $\phi = v$ corresponds to the minimum of the function $|f(\phi)^n|$, therefore the ground state of the potential $V_{J}(\phi) \propto f(\phi)^2$. Finally, $\phi_c$ is the largest field value where one can approximately take that $f(\phi_i) \simeq f(\phi_c)$.
}
\end{figure}

Here for the Jordan frame TCC (\ref{NMTCC}), we claim that; it is also logical to consider the effective Planck mass $\mathcal{M}_{\mathrm{pl}}(\phi_i)$ with its initial field value as a constant. That should not be understood as a modification of the Jordan frame TCC, but only an approximate implementation of a model which will be exhibited in the continuation of this subsection.\footnote{See also \cite{Shi:2020ymp} for a similar argument about considering the initial value of the effective Planck mass, together with an asymptotically safe period scenario at the early universe.} According to this, one gets the Jordan frame TCC conjecture as
\begin{equation}\label{NMTCCi}
    a_{f}  H_{f} < a_{i} \mathcal{M}_{\mathrm{pl}}(\phi_i)
\end{equation}
and following this, the bound on Jordan frame potential reads:
\begin{equation}\label{VJbound}
     V_{J}(\phi) \lesssim  10^{-40} \left( M_{\mathrm{pl}}^2 + \xi f(\phi_i) \right)^2.
\end{equation}
One can easily see that the upper bound on the Jordan frame potential eases at a strong coupling limit as $ |\xi f(\phi_i)| \gg M_{\mathrm{pl}}^2$. As we continue, let us investigate the Einstein frame slow-roll parameter, $\epsilon_{E}$ in more detail for this claim. It is possible to decompose this term as denoted in \cite{Linde:2011nh}: 
\begin{equation}\label{Zrelation}
\epsilon_{E}(\varphi) = Z(\phi) \epsilon_{\text{minimal}}(\phi)
\end{equation}
where $\epsilon_{\text{minimal}}$ is same the as (\ref{epsilonminimal}). One can immediately see that for a constant non-minimal coupling with a fixed value of $f\left(\phi_{i}\right) \simeq f\left(\phi_{c}\right)$, the non-minimal coupling term becomes $F(\phi) \rightarrow F(\phi_{c}) : \text{constant}$. Then one gets $Z(\phi_c) = F(\phi_c)/M_{\mathrm{pl}}^2$. This manifests a linear relation between fields via the definition of the canonical scalar field $\varphi$  as
    \begin{equation}
         \varphi = \pm \frac{M_{\mathrm{pl}}}{\sqrt{F(\phi_c)}} \phi.
    \end{equation}
     Using the same relation (\ref{Zrelation}) to determine the slow-roll parameter $\epsilon$ in the Einstein frame, one gets:
\begin{equation}
    \epsilon_{E}(\varphi) =   \frac{F(\phi_c)}{M_{\mathrm{pl}}^2} \epsilon_{\text{minimal}}(\phi).
\end{equation}
One can also investigate this case for the Jordan frame with using the modified gravitational sector approach by setting $M_{\mathrm{pl}}^2 \rightarrow \mathcal{M}_{\mathrm{pl}}(\phi_c)^2 = F(\phi_c)$, and gets:
\begin{equation}\label{epsilonJconst}
    \epsilon_{J}(\phi) = \frac{\mathcal{M}_{\mathrm{pl}}(\phi_c)^2}{2} \left(\frac{V^{\prime}(\phi)}{V(\phi)}\right)^{2} = \frac{F(\phi_c)}{M_{\mathrm{pl}}^2} \epsilon_{\text{minimal}}(\phi),
\end{equation}
therefore one can denote that $\epsilon = \epsilon_{J} = \epsilon_{E}$ for a constant effective Planck mass, as expected. There is a perfect inflationary model to see the further physical consequences of this claim, which is the \textit{new inflation} model as proposed in \cite{Albrecht:1982wi,Linde:1981mu}. During the inflation era, the scalar field in this model; very slowly rolls from its symmetric state $\phi=0$, into the ground (symmetry-breaking) state $\phi=v$, together with a potential that includes a vacuum expectation value $v$. This potential gets very flat in order to allow slow-rolling to provide enough e-folds and behaves as almost like a constant in a long period during inflation. In that sense, one can identify the scalar potential as 
\begin{equation}
    V_{J}(\phi)=\tau^2 f(\phi)^2
\end{equation}
to be compatible with the generic non-minimal coupling term
\begin{equation}
    \mathcal{M}_{\mathrm{pl}}(\phi)^2 \equiv F(\phi) = M_{\mathrm{pl}}^2 + \xi f(\phi).
\end{equation}
This is called \emph{the generalized non-minimal coupling}. Here, we only assume that the scalar potential is positive, and it is in the Jordan frame because of the existence of the non-minimal coupling term. For this case, effective Planck mass is directly correlated with the potential term considering the relation $  \mathcal{M}_{\mathrm{pl}}(\phi) \propto V_{J}(\phi)^{1/4}$. Therefore, generalized non-minimal coupling (Jordan frame + Model) points to a model-dependent scenario. Consequently, it is safe to say that there is a long period that the relation $|f(\phi_i)| \simeq |f(\phi_c)| \gtrsim |f(\phi_f)| $ holds for the generalized non-minimal coupling case with a flat potential; see Figure \ref{F1}.
Therefore, it is a good approach to treat the effective Planck mass as a constant for a large period of time of the new inflation model. Hence, the slow-roll parameter
\begin{equation}\label{ConstantEp}
    \epsilon(\phi) =  \frac{F(\phi_i)}{M_{\mathrm{pl}}^2} \epsilon_{\text{minimal}}(\phi) \simeq \frac{F(\phi_c)}{M_{\mathrm{pl}}^2} \epsilon_{\text{minimal}}(\phi) :=  \epsilon_c(\phi)
\end{equation}
of the constant non-minimal coupling case has a physical meaning in this period. Here we define this parameter as $\epsilon_c(\phi)$, to represent the slow-roll parameter for the constant non-minimal coupling case between the field values of $\phi_i$ and $\phi_c$. Then, using (\ref{Zrelation}), one can denote that:
\begin{equation}\label{EpsilonC1}
    \begin{aligned}
        \epsilon_c(\phi) &=\frac{F(\phi_i)}{M_{\mathrm{pl}}^2} Z(\phi)^{-1} \epsilon_E (\varphi) \\
        &=F(\phi_i) \left[\frac{1}{ F(\phi)} +\frac{3}{2 }\left(\log(F(\phi))\right)^{\prime 2}\right] \epsilon_E (\varphi) .
    \end{aligned}
\end{equation}
For the generalized non-minimal coupling term $F(\phi)=M_{\mathrm{pl}}^2 + \xi f(\phi)$ with the general inflation potential $V(\phi)=\tau^2 f(\phi)^2$, the equation (\ref{EpsilonC1}) can more explicitly be denoted as:
\begin{equation}\label{EpsilonC2}
     \epsilon_c(\phi) =(M_{\mathrm{pl}}^2 + \xi f(\phi_i)) \left[\frac{1}{M_{\mathrm{pl}}^2 + \xi f(\phi)} +\frac{3}{2 }\left( \frac{\xi f(\phi)^{\prime}}{M_{\mathrm{pl}}^2 + \xi f(\phi)} \right)^2\right] \epsilon_E (\varphi). 
\end{equation}
Using this, we will investigate the weak and strong coupling limits by following the same approximations with the ref. \cite{Kallosh:2013tua}. 
\begin{enumerate}
    \item[\textbf{i.}] {\textbf{Weak coupling:}}
     For small values of the non-minimal coupling term, i.e. $|\xi f(\phi)| \ll M_{\mathrm{pl}}^2$, suppressing the higher-order terms in $Z(\phi)$, one gets the relation between fields as
    \begin{equation}
        \frac{d \varphi}{d \phi} \simeq \pm \left( 1 - \frac{\xi}{2 M_{\mathrm{pl}}^2} f(\phi) \right)
    \end{equation}
and the Einstein frame potential as
\begin{equation}
    V_{E}(\varphi) \simeq \tau^2 f(\phi)^2 \left( 1- \frac{2 \xi f(\phi)}{M_{\mathrm{pl}}^2} \right).
\end{equation}
For this case, the general slow-roll parameter equation (\ref{EpsilonC2}) leads to:
\begin{equation}
    \begin{aligned}
    \epsilon_c(\phi) &\simeq M_{\mathrm{pl}}^2 \left[\frac{1}{M_{\mathrm{pl}}^2 } +\frac{3}{2 }\left( \frac{\xi f(\phi)^{\prime}}{M_{\mathrm{pl}}^2 } \right)^2\right] \epsilon_E (\varphi) \\
    &\simeq  \left[1 +3\left( \frac{\xi f(\phi)}{M_{\mathrm{pl}}^2 } \right)^2 \epsilon_c (\phi) \right] \epsilon_E (\varphi) \\ &\simeq \epsilon_E (\varphi),
    \end{aligned}
\end{equation}
and the upper bound on $r$ remains unchanged for the constant non-minimal coupling era. Here, we also used the equation (\ref{ConstantEp}) and consider the weak coupling approximation as $|\xi f(\phi_i)| \ll M_{\mathrm{pl}}^2$ including it's initial value. 

    \item[\textbf{ii.}] {\textbf{Strong coupling:}}
    For this case, the limit $|\xi f(\phi)| \gg M_{\mathrm{pl}}^2$ leads to the canonical field $\varphi$:
    \begin{equation}
        \varphi \simeq \pm \sqrt{\frac{3}{2}} M_{\mathrm{pl}} \log(F(\phi)),
    \end{equation}
    where we assume that $Z(\phi)$ gets contribution only from higher-order terms. Then, the Einstein frame potential becomes:
    \begin{equation}
        V_{E}(\varphi) \simeq \frac{  \tau^{2}}{\xi^{2}} M_{\mathrm{pl}}^4 \left[1-\exp{\left(-\sqrt{\frac{2}{3}} \frac{\varphi}{M_{\mathrm{pl}}}\right)}\right]^{2}
    \end{equation}
    in case $f(\phi)$ is odd in $\phi$. The Einstein frame Lagrangian (\ref{EL1}) corresponds to the scalar form of the $R+R^2$ (Starobisnky) model \cite{Starobinsky:1980te} for this case. If $f(\phi)$ is even in $\phi$ then the Einstein frame potential becomes: 
    \begin{equation}
        V_{E}(\varphi) \simeq \frac{  \tau^{2}}{\xi^{2}} M_{\mathrm{pl}}^4 \left[1-\exp{\left(-\sqrt{\frac{2 \varphi^2 }{3 M_{\mathrm{pl}}^2}} \right)}\right]^{2},
    \end{equation}
    and for this case, one gets the symmetry $\varphi \rightarrow -\varphi$ for the canonical field $\varphi$. Here, one should note that; we are completely disconnected with the original Jordan frame potential $ V_J(\phi) = \tau^2 f(\phi)^2 $ except  its coupling constant parameter $\tau^2$. This is why the generalized non-minimal coupling term leads to a universal attractor behavior \cite{Kallosh:2013tua}.
    As one continues for the strong coupling case with $|\xi f(\phi)| \gg M_{\mathrm{pl}}^2$, the slow-roll parameter (\ref{EpsilonC2}) reads\footnote{Note that we also consider higher-order terms here.}:
    \begin{equation}
        \begin{aligned}
         \epsilon_c(\phi) &\simeq \xi f(\phi_i) \left[\frac{1}{\xi f(\phi)} +\frac{3}{2 }\left( \frac{ f(\phi)^{\prime}}{  f(\phi)} \right)^2\right] \epsilon_E (\varphi) \\
         &\simeq  \left[\frac{f(\phi_i)}{ f(\phi)} +3 \epsilon_{c}(\phi) \right] \epsilon_E (\varphi) \\
         &=\left[\frac{f(\phi_i)}{ f(\phi) \left(1-3 \epsilon_E (\varphi) \right)}\right] \epsilon_E (\varphi).
         \end{aligned}
    \end{equation}
    Again here, we used the equation (\ref{ConstantEp}) and allow the strong coupling approximation as $|\xi f(\phi_i)| \gg M_{\mathrm{pl}}^2$ including it's initial value. At this point, using the inequality (\ref{rEbound}), one can denote the bound on the parameter $r$ for the constant non-minimal coupling era as:
    \begin{equation}\label{StrongrBound}
        r = 16 \epsilon_c < \frac{f(\phi_i)}{ f(\phi)} 10^{-30}.
    \end{equation}
    This inequality implies that the bound on $r$ is independent of the non-minimal coupling constant $\xi$, or the coupling constant of the potential, $\tau^2$. One can also see that the bound on $r$ eases for $f(\phi_i) \simeq f(\phi) \times 10^{29}$ and becomes compatible with the latest observational expectations \cite{Akrami:2018odb}, i.e. $r \lesssim 10^{-1}$. It is clearly seen in Figure \ref{F1} that; this situation will be provided naturally as the value of $f(\phi)$ gets closer to zero, and hence when the field value of $\phi$ approaches the symmetry-breaking point $\phi=v$ for the new inflation models.
\end{enumerate}
Although the example we give here is for the new inflation models with symmetry-breaking type of potentials, one can consider the equation (\ref{StrongrBound}) more generally. In this case, the only precondition is that the slow-roll parameter $\epsilon$ is to be physically meaningful for the initial value of the non-minimal coupling term as it was in our original claim, (\ref{NMTCCi}). Because the original TCC has been formulated in the Einstein frame, and the choice of the Jordan frame does not affect it. Yet, the choice of the Jordan frame affects the resulting inequality (\ref{StrongrBound}) for the TCC in the Jordan frame. For example, instead of using the generalized non-minimal coupling (with a Higgs-like potential)
\begin{equation}
    \mathcal{M}_{\mathrm{pl}}(\phi)^2= M_{\mathrm{pl}}^2 + \xi f(\phi) \quad \text{with} \quad V_{J}(\phi)=\tau^2 f(\phi)^2,
\end{equation}
one can choose a non-minimal coupling term, or equivalently an effective Planck mass, as
\begin{equation}
    \mathcal{M}_{\mathrm{pl}}(\phi)^2 = M_{\mathrm{pl}}^2 + \xi g(\phi)
\end{equation}
where $g(\phi) = f(\phi)^{-1}$, inversely correlated with the same Jordan frame potential. Then the resulting inequality (at strong coupling limit) would be even more constraining in the Jordan frame as
\begin{equation}
     r  \lesssim \frac{f(\phi)}{ f(\phi_i)} 10^{-30}.
\end{equation}
Here, when the value of $f(\phi)$ gets closer to zero, or the potential term reaches the ground state at the symmetry-breaking value of inflaton $\phi = v$, the constraint on the tensor-to-scalar ratio gets tighter than the original Einstein frame constraints of the TCC. Therefore, if one desires to relax the TCC upper bounds on inflationary cosmology, it is clear to see that; TCC in the Jordan frame estimates a model-dependent selection rule (or cutoff) on the possible infinite Jordan frames in that perspective.

\subsection{Superconformal Embedding \texorpdfstring{($\mathcal{N}=1$, $D=4$)}{TEXT}}\label{sec3.2}
Introducing additional functional freedom on the Lagrangian via adding a generalized non-minimal coupling term served us well through removing the TCC bound on the observational parameter $r$ and easing the TCC bound on the Jordan frame potential $V_J(\phi)$. Since we are dealing with a string swampland conjecture, it is natural to construct our frame with string theory motivations. In this sense, one can employ supergravity that acts as a bridge to the supersymmetric string theory \cite{Green:1984sg,Bergshoeff:1987cm,Hull:1994ys,Witten:1995ex}. (See also \cite{VanProeyen:2020uof} for a chronological list of developments and a compact review.) Then, this non-minimal coupling term can be embedded in $\mathcal{N}=1$, $D=4$ supergravity. In a nutshell, the embedding process generally follows the steps below.

First, one needs to build a generic F-term supergravity potential by choosing specific symmetries on the Kähler potential. String theory motivations and (nilpotent) supergravity \cite{Dudas:2015eha,Ferrara:2015tyn,Kallosh:2014via,Ferrara:2014kva} correspondences of this type of Kähler potentials that lead to generic potentials are argued and demonstrated in refs. \cite{Ferrara:2014kva,Roest:2013aoa}. Following \cite{Kallosh:2010xz}, these symmetries are given as
\begin{equation}
    \begin{aligned}
&S \rightarrow-S \\
&\Phi \rightarrow \bar{\Phi}  \\
&\Phi \rightarrow \Phi+a
    \end{aligned}
\end{equation}
where  $a \in \mathbb{R} $. Here $\Phi$ and $S$ are scalar fields of chiral multiplets. The field $\Phi$ contains inflaton field $\phi$ and the field $S$ is usually considered as a goldstino supermultiplet which also serves as a stabilization term \cite{Kallosh:2010xz,Ferrara:2010in,Lee:2010hj,Ferrara:2010yw}. The resulting Kähler potentials lead to generic F-term supergravity potentials as 
\begin{equation}
    V_F (\Phi) \propto \pm  f(\Phi)^2
\end{equation}
with a superpotential term as 
\begin{equation}
   W(S,\Phi) \propto S f(\Phi).
\end{equation}
Here, one considers the inflationary trajectory where all fields disappear except the inflaton field. For example, one can decompose these fields as 
\begin{equation}
S=\frac{1}{\sqrt{2}}(s+i \lambda), \quad \Phi=\frac{1}{\sqrt{2}}(\phi+i \beta)
\end{equation}
where $s,\lambda,\phi,\beta$ are canonically normalized real fields. For this representation, the inflationary trajectory reads as $s=\lambda=\beta=0$ or $S=\Phi -\bar{\Phi} =0$.

The next step is to insert additional terms to the corresponding (superconformal) Kähler potential. It is appropriate to consider the relevant scalar-gravity part 
\begin{equation}
  \mathcal{L}_{\text{sc}}^{\text{scalar-gravity}}=  \sqrt{-g}\left[-\frac{1}{6} N R \right] +  \mathcal{L}_{\text{Kinetic Terms}} +  \mathcal{L}_{\text{Potential Terms}}
\end{equation}
of the superconformal Lagrangian as given in \cite{Ferrara:2010in} for this step. In this notation, the (reduced) Planck mass is taken as $M_{\mathrm{pl}} \equiv 1$. So, as a fitting superconformal Kähler potential for this task, one can generally denote that
\begin{equation}
    N(S, \bar{S},\Phi,\bar{\Phi})' \rightarrow N(S, \bar{S},\Phi,\bar{\Phi}) + \alpha \left(f\left(\frac{\Phi}{\alpha'}\right) + f\left(\bar{\frac{\Phi}{\alpha'}}\right) \right)
\end{equation}
where $\alpha$($\alpha'$) contains the parameters of the theory, which is also necessary for properly adjusting the Weyl weights, and $N$ is the superconformal Kähler potential. These additional terms do not disturb the generality of the F-term potential. Equivalently, the same procedure can also be done without lifting our theory to the superconformal level by adding logarithmic terms $\log\left[\alpha \left(f(\Phi) + f(\bar{\Phi}) \right)\right]$ to the Kähler potential itself as denoted in \cite{Kallosh:2013tua}.

Finally, after fixing special conformal symmetries and dilation symmetries appropriately, one can reach the Jordan frame inflationary  Lagrangian of the form (\ref{JL1}) with $V_J(\phi) = \tau^2 f(\phi)^2 $, at the inflationary trajectory as
\begin{equation}
   \mathcal{L}_J =\sqrt{-g}\left[\frac{R}{2} +\xi f(\Phi) R - \left(\partial \Phi \right)^2-\tau^2 f(\Phi)^2\right].
\end{equation}
Also, one should note that the sign of the supergravity F-term potential is not fixed and is determined by the theory. As shown in \cite{Atli:2020dni}, even with a negative signed effective potential term, one can configure the setup accordingly to obtain a Lagrangian that corresponds to a de Sitter solution of gravity, such as inflation theory.

\section{Conclusions}\label{sec4}
In this work, firstly, we demonstrated a general embedding of the generic non-minimal coupling term to the TCC for both Einstein and Jordan frames. By doing so, we denoted that both frames equivalently lead to the same upper bound on the inflationary parameter $r$. But we also denoted that the bound on the Jordan frame potential $V_J(\phi)$ could ease at strong coupling limit of the generic non-minimal coupling term as one can see from the inequality
\begin{equation}
    V_{J}(\phi) \lesssim 10^{-40}  F(\phi)^2 \quad \text{or} \quad V_{J}(\phi) \lesssim 10^{-40}  \mathcal{M}_{\mathrm{pl}}(\phi)^4.
\end{equation}
Where here $F(\phi) = M_{\mathrm{pl}}^2 + \xi f(\phi)$ is a generic non-minimal coupling term, with a generic function $f(\phi)$.  Equivalently to that, $\mathcal{M}_{\mathrm{pl}}(\phi)$ is the effective (reduced) Planck mass. Furthermore, we presented an example for the generalized non-minimal coupling case. For that, we assumed the TCC as
\begin{equation}
    a_{f}  H_{f} < a_{i} \mathcal{M}_{\mathrm{pl}}(\phi_i),
\end{equation}
with the initial value of the effective Planck mass as a constant. This approach allowed us to calculate the TCC bound on $r$, as
\begin{equation}
        r  < \frac{f(\phi_i)}{ f(\phi)} 10^{-30}
\end{equation}
at the strong coupling limit of the generalized non-minimal coupling case, that leads to a universal attractor behavior independently of the original Jordan frame potential, $V_J$. This inequality also showed that; when the scalar potential of the theory, $V_J(\phi)= \tau^2 f(\phi)^2$, reaches the symmetry-breaking point at $\phi = v$, that upper bound naturally disappears for Higgs-like symmetry breaking potentials. At this point, the term $f(\phi)$ disappears and as a consequence, one gets the usual Planck mass $M_{\mathrm{pl}}$.
\acknowledgments

The author thanks Mehmet Ozkan for his valuable suggestions.


\end{document}